\documentclass[a4paper]{PoS}
\usepackage{physics}
\usepackage{mathrsfs}
\usepackage{subcaption}

\newcommand{\beq}{\begin{equation}}
\newcommand{\eeq}{\end{equation}}

\title{Scattering phase shift determinations from a two-scalar field theory}

\ShortTitle{Scattering phase shift determinations from a two-scalar field theory}

\author{\speaker{Daniel Darvish}$\;^a$,Ruair\'i Brett$^a$, John Bulava$^b$, Jacob Fallica$^c$,
  Andrew Hanlon$^d$,Colin Morningstar$^a$\\
  \llap{$^a$}Dept. of Physics, Carnegie Mellon University,
  Pittsburgh, PA 15213, USA\\
  \llap{$^b$}Dept. of Mathematics and Computer Science and CP3-Origins,
  University of Southern Denmark,
  Campusvej 55, 5230 Odense M, Denmark\\
  \llap{$^c$}Department of Physics and Astronomy, University of Kentucky,
  Lexington, KY 40506, USA\\
  \llap{$^d$}Helmholtz-Institut Mainz, Johannes Gutenberg-Universit\"at,
  55099 Mainz, Germany\\
  E-mail: \email{darvish@cmu.edu}}

\abstract{A field theory involving two interacting scalar fields, previously studied
by Rummukainen and Gottlieb, is revisited.   Our study is not restricted to the limit of large 
quartic couplings, and a Symanzik-improved action is used so that continuum dispersion relations
work well.  The Metropolis method, combined with a local microcanonical updating algorithm,
is employed in our Monte Carlo calculations.  Isotropic lattices ranging from $16^3 \times 48$ to $
53^3 \times 48$ are used, and scattering phase shifts are determined using a L\"uscher analysis
with multiple partial waves.}

\FullConference{The 36th Annual International Symposium on Lattice Field Theory - LATTICE2018\\
		22-28 July, 2018\\
		Michigan State University, East Lansing, Michigan, USA.}

\begin{document}

\section{Introduction}
A theory of two real scalar fields, $\phi$ and $\rho$, described by the Lagrangian,
\beq \label{eq:L}
\mathscr L = -\frac{1}{2}\phi(\partial^2 + M_\phi^2)\phi -\frac{1}{2}\rho(\partial^2 + M_\rho^2)\rho 
- \frac{\lambda_\phi}{4!}\phi^4 - \frac{\lambda_\rho}{4!}\rho^4 - \frac{1}{2}g\phi^2\rho ,
\eeq
is studied here.  Parameters in the above Lagrangian are chosen such that the $\rho$ particle may 
decay into two $\phi$ particles in infinite volume.  We also ensure that spontaneous symmetry breaking
does not occur. We formulate this theory on a space-time lattice and 
study $\phi$-$\phi$ scattering to extract resonance parameters of the theory using the box matrix 
formalism developed in Ref.~\cite{Morningstar2017}, which is an implementation of
the so-called L\"uscher method\cite{Luscher1991,Kim2005}.   This model was previously studied in 
1995 by Rummukainen and Gottlieb~\cite{Rummukainen1995} in the Ising limit.  Here, we do 
not take the large limit in the quartic self-couplings, and we avoid using unphysical lattice 
dispersion relations by making use of a tree-level Symanzik-improved action.

\section{Choice of model parameters}
We require the physical $\rho$ mass to be greater than twice the physical $\phi$ mass, so that the decay 
is kinematically allowed, and less than three times the $\phi$ mass, as required for the 
L\"uscher analysis. Quartic interactions in Eq.~(\ref{eq:L}) are included so that our action is bounded 
from below in Euclidean space-time, but we wish to keep these couplings as small as possible to reduce 
mass renormalizations. We also require that our theory features no spontaneous symmetry breaking, so we 
design the action to have a unique minimum at $\phi=\rho=0$. To reduce finite-volume effects, we impose the 
condition $m_\phi L > 4$, where $m_\phi$ is the measured $\phi$ mass and $L$ is the spatial extent of the lattice. Then we choose $a_t M_\phi$ according to this constraint and our choice of lattice size, where $a_t$ is the temporal lattice 
spacing. Finally, we need to pick a value for the 1-to-2 coupling, $g$, large enough to produce 
significant interaction energies. Table~\ref{table:params} gives the run parameters we chose to 
satisfy the above conditions.

\begin{table}[h]
\centering
\begin{tabular}{c c c c c}
\hline
$a_t M_\phi$ &
$a_t M_\rho$ &
$\frac{g}{M_\phi}$ &
$\lambda_\phi$ &  
$\lambda_\rho$ \\
\hline
$0.1$ & $0.31$ & $1$ & $\frac{g^2}{4 M_\phi^2}$ &  $\frac{g^2}{M_\phi^2}$\\
\hline
\end{tabular}
\caption{Model parameters}\label{table:params}
\end{table}

\section{The need for an improved action}
\begin{figure}
\centering
\includegraphics[scale=.35]{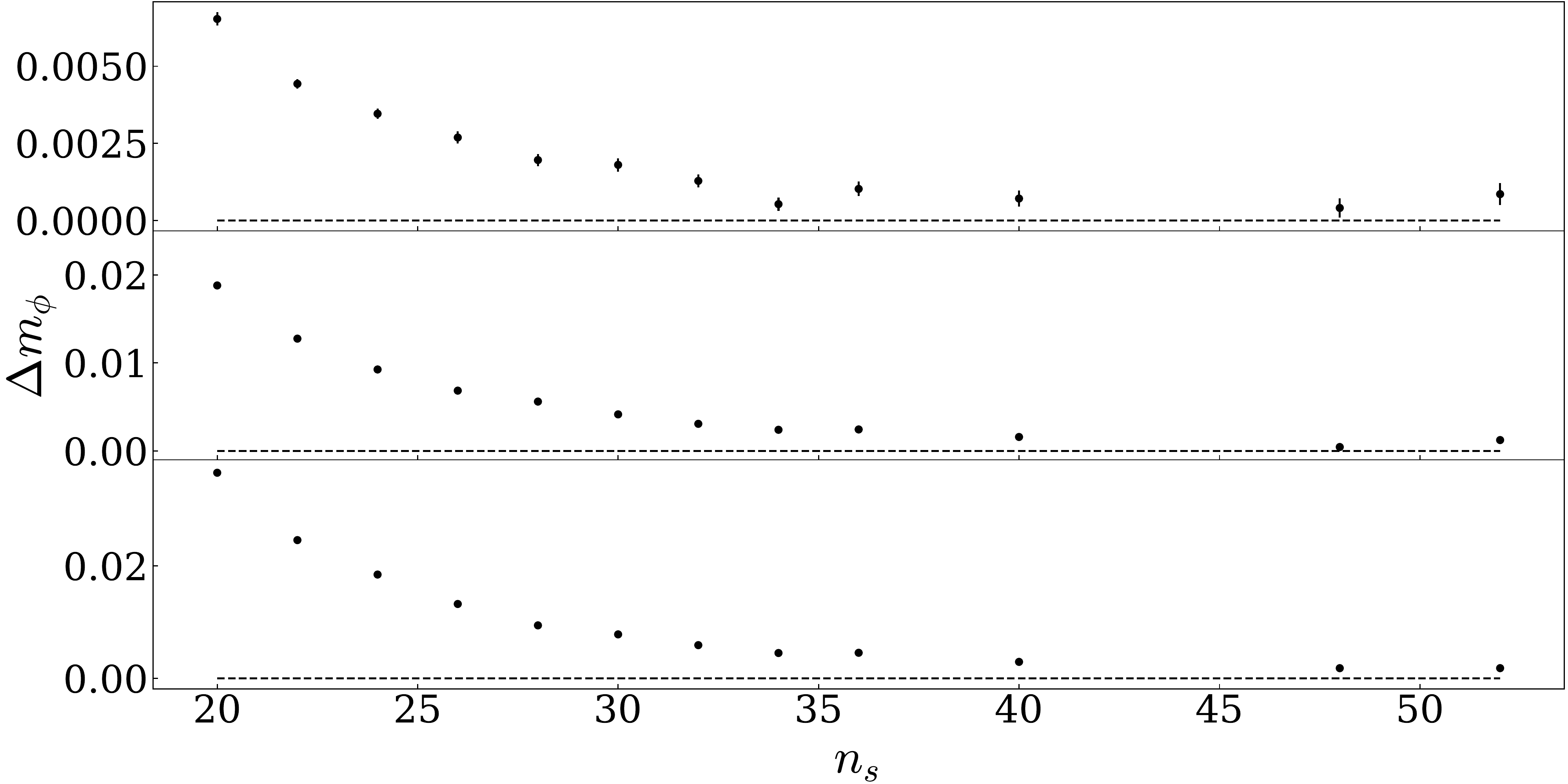}
\caption{Differences between moving-frame and rest-frame measurements of the $\phi$ mass, using the unimproved action.}\label{fig:delta_m_unimp}
\end{figure}

If we use a naive discretization scheme involving only finite differences in the derivative terms of 
the action, then the portion of the action including only the $\phi$ field can be written as,
\beq\label{eq:unimp_action}
S_\phi = a_s^{D-1}a_t\sum_x\sum_\mu \left(\frac{(\phi(x+a_\mu \hat\mu)-\phi(x))^2}{2a_\mu} 
+ \frac{1}{2}M_\phi^2 \phi(x)^2  + \frac{\lambda_\phi}{4!}\phi(x)^4\right),
\eeq
where $a_s$ denotes the spatial lattice spacing. In our work, we use an isotropic lattice. We 
find that using this scheme produces overly-large discretization errors when performing boosts 
with the continuum dispersion relation, as shown in Fig.~\ref{fig:delta_m_unimp}.
Let $E_\phi^{\vb{P}}$ denote the energy of the $\phi$ particle having momentum 
$\vb{P}=\frac{2\pi\vb{d}}{L}$, where $\vb{d}$ is a vector of integers and $L=a_s n_s$ is the 
length of the isotropic $L^3$ lattice.  We determine $E_\phi^{\vb{P}}$ from the exponential fall-off
of appropriate temporal correlation functions estimated by the Monte Carlo method in the
standard way.  The difference between the $\phi$ mass measured in the moving frame 
and its mass in the rest frame is
$\Delta m_\phi=[(E_\phi^{\vb{P}})^2-\vb{P}^2]^{1/2}-E_\phi^{\vb{0}}$.  In the continuum limit,
this quantity should vanish.  Any deviation from zero is a measure of discretization errors.
One sees sizeable discretization errors in Fig.~\ref{fig:delta_m_unimp}.

\begin{figure}
\centering
\includegraphics[scale=.35]{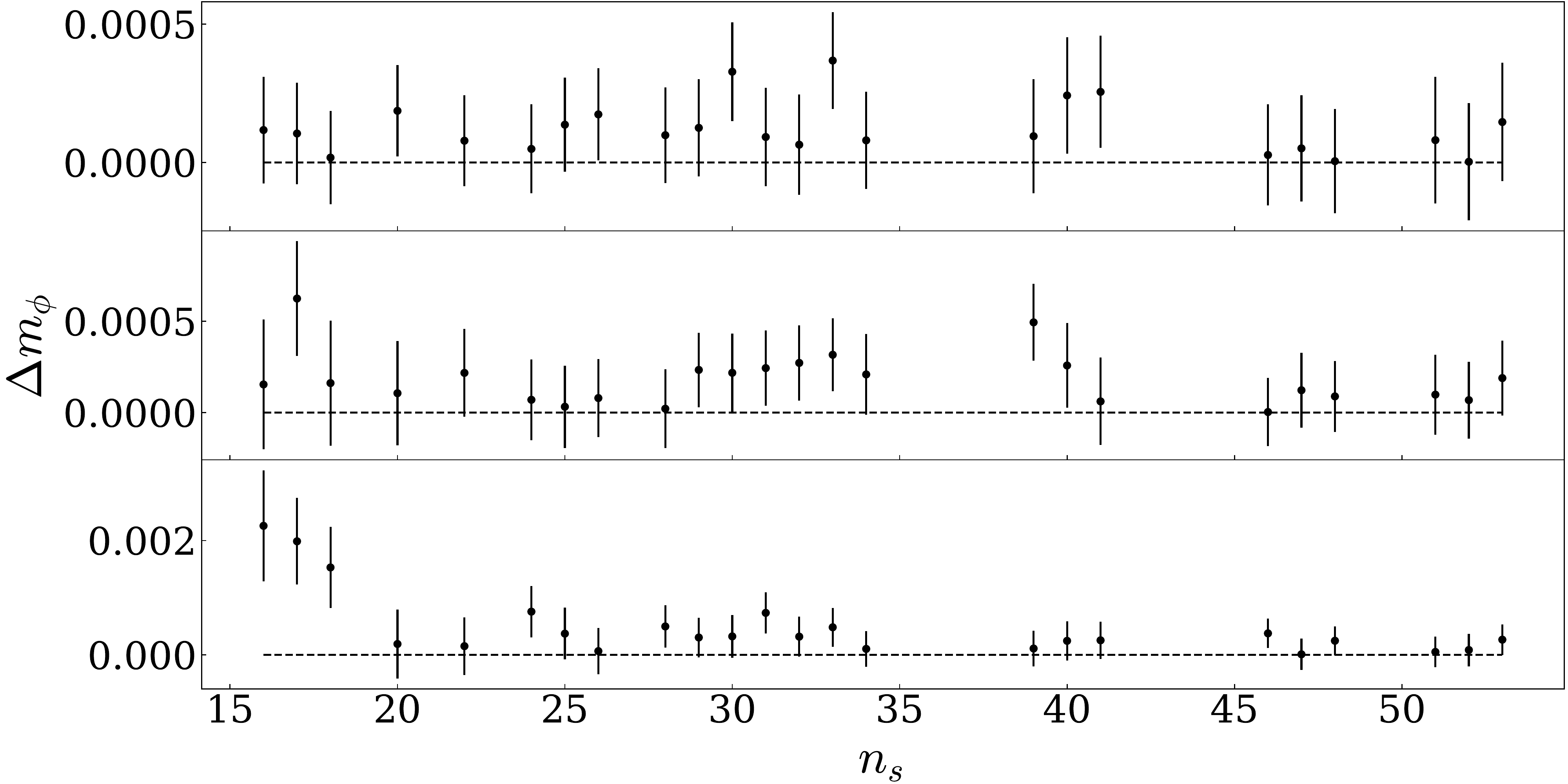}
\caption{Differences between moving-frame and rest-frame measurements of the $\phi$ mass, using the Symanzik improved action.}\label{fig:delta_m_imp}
\end{figure}

Ref.~\cite{Rummukainen1995} handled this problem by making use 
of ``lattice'' dispersion relations.  We, instead, decided to employ a tree-level
Symanzik-improved action and found that with such an action, continuum energy-momentum
dispersion relations worked well.
For a lattice spacing $a_s$, the discretization error in the finite-difference approximation of the action 
is $\mathcal O(a_\mu^2)$. We can reduce this error down to $\mathcal O(a_\mu^4)$ by employing a 
tree-level Symanzik improvement, which is given below for the portion of the action containing only the 
$\phi$ terms (the $\rho$ terms are similar):
\beq
\begin{aligned}
S_\phi^I &=  a_s^{D-1}a_t\sum_x  \left\lbrace  \frac{1}{2a_\mu^2}\sum_\mu
\left(-\frac{4}{3}\phi(x+a_\mu)\phi(x)-\frac{4}{3}\phi(x-a_\mu)\phi(x)\right. \right. \\
&\left. +\frac{1}{12}\phi(x+2a_\mu)\phi(x) + \frac{1}{12}\phi(x-2a_\mu)\phi(x) 
+ \frac{5}{2}\phi(x)^2\right) \\
&\left. +\frac{1}{2} M_\phi^2 \phi(x)^2 + \frac{\lambda_\phi}{4!}\phi(x)^4\right\rbrace.
\end{aligned}
\eeq
Fig.~\ref{fig:delta_m_imp} shows $\Delta m_\phi$ using the Symanzik improved action, and demonstrates 
the ameliorative effect that the improved action has on the discretization errors in the boosts.


\section{Phase shift determination and resonance parameter extraction}
\begin{figure}[h]
\centering
\includegraphics[height=2in]{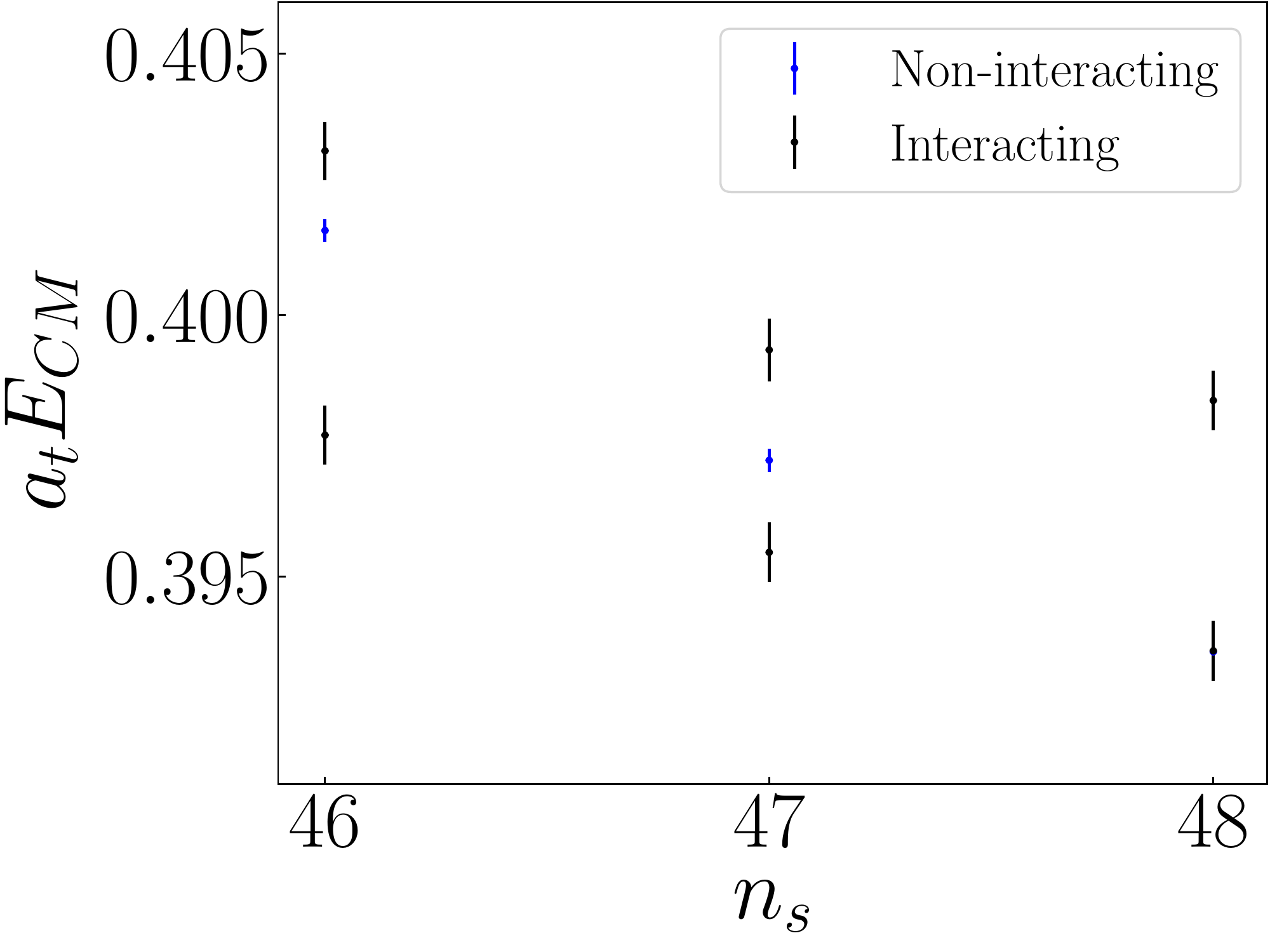}
\includegraphics[height=2in]{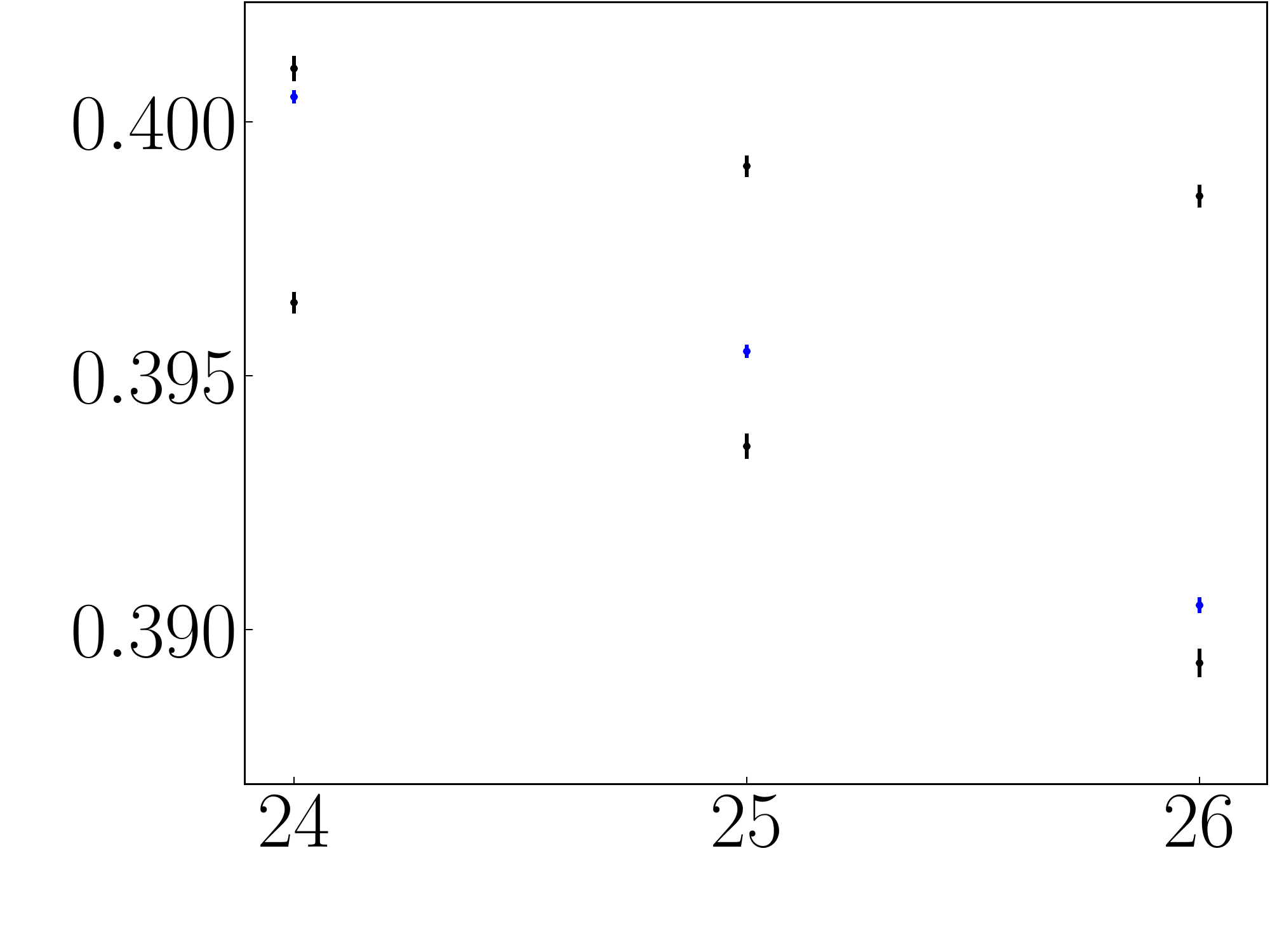}
\caption{Center-of-momentum energies near avoided level crossings in the (left) $\vb{d}^2=0$ channel
and the (right) $\vb{d}^2=2$ channel.\label{fig:avoided_0}}
\end{figure}

\begin{figure}[h]
	\centering
	\begin{subfigure}[h]{0.49\textwidth}
		\centering
		\includegraphics[height=2in]{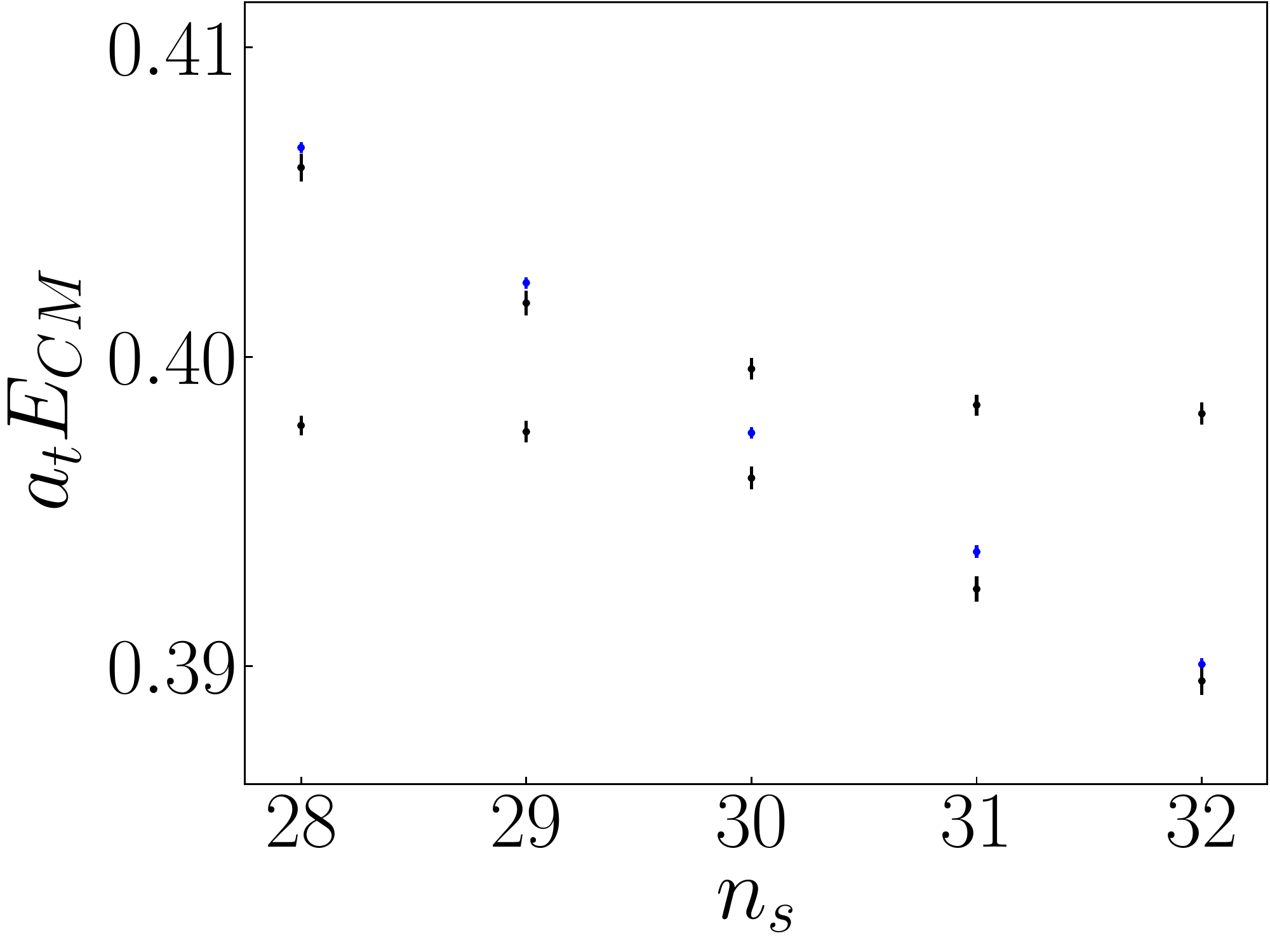}
	\end{subfigure}
	\begin{subfigure}[h]{0.49\textwidth}
		\centering
		\includegraphics[height=2in]{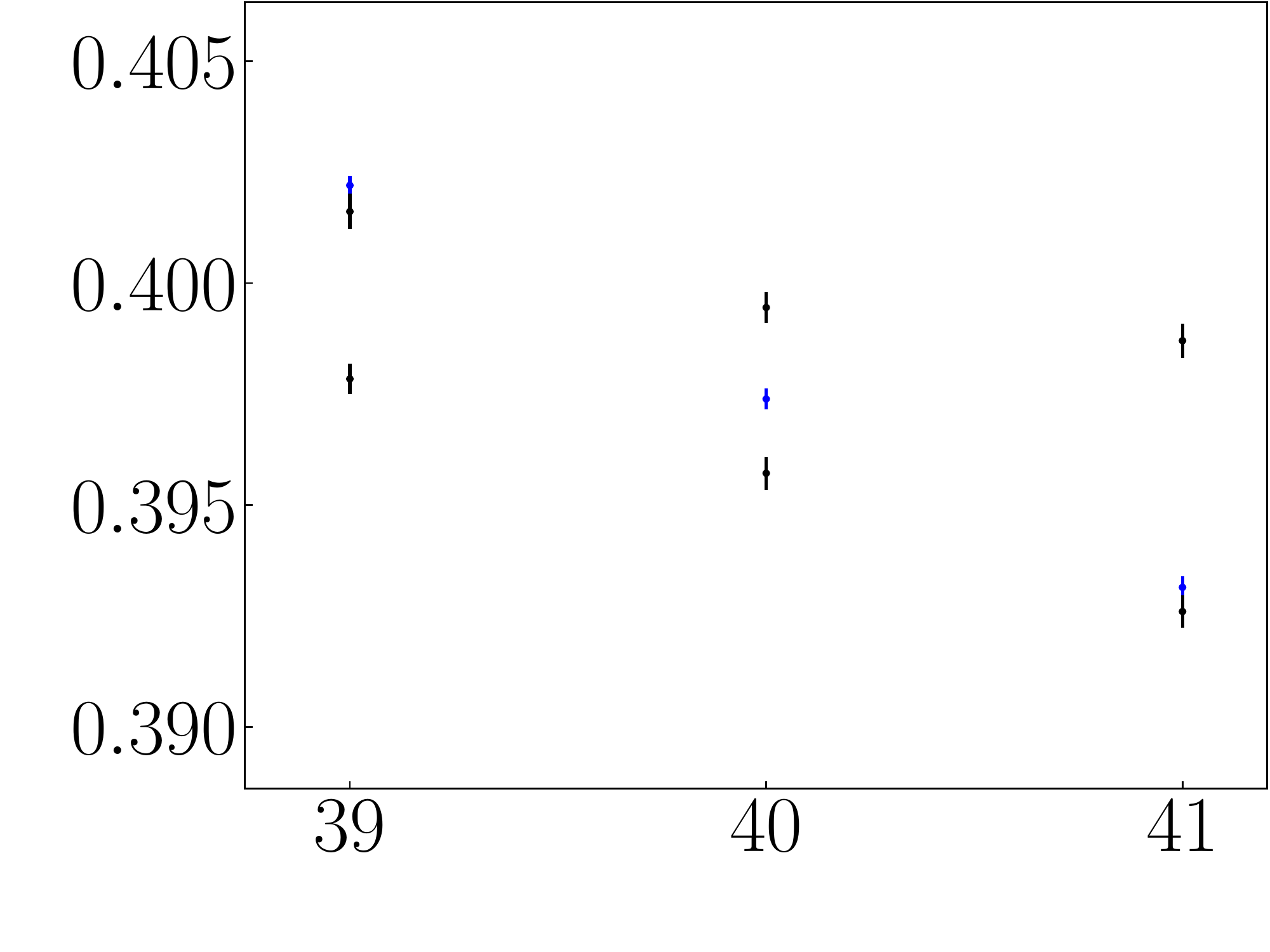}
	\end{subfigure}
\caption{Avoided level crossings in the $\vb{d}^2=3$ channel, boosted to the center-of-momentum frame.}\label{fig:avoided_3}
\end{figure}

A large portion of the finite-volume spectrum was determined, but for the purpose of the L\"uscher analysis, 
we are only interested in energies which differ significantly from their noninteracting values and are below the $3m_\phi$ threshold. These energies manifest as avoided level crossings, and Figs.~\ref{fig:avoided_0}-\ref{fig:avoided_3} show 
examples of these avoided level crossings in the center-of-momentum frame.

\begin{figure}[h]
	\centering
	\begin{subfigure}[h]{0.49\textwidth}
		\centering
		\includegraphics[height=2in]{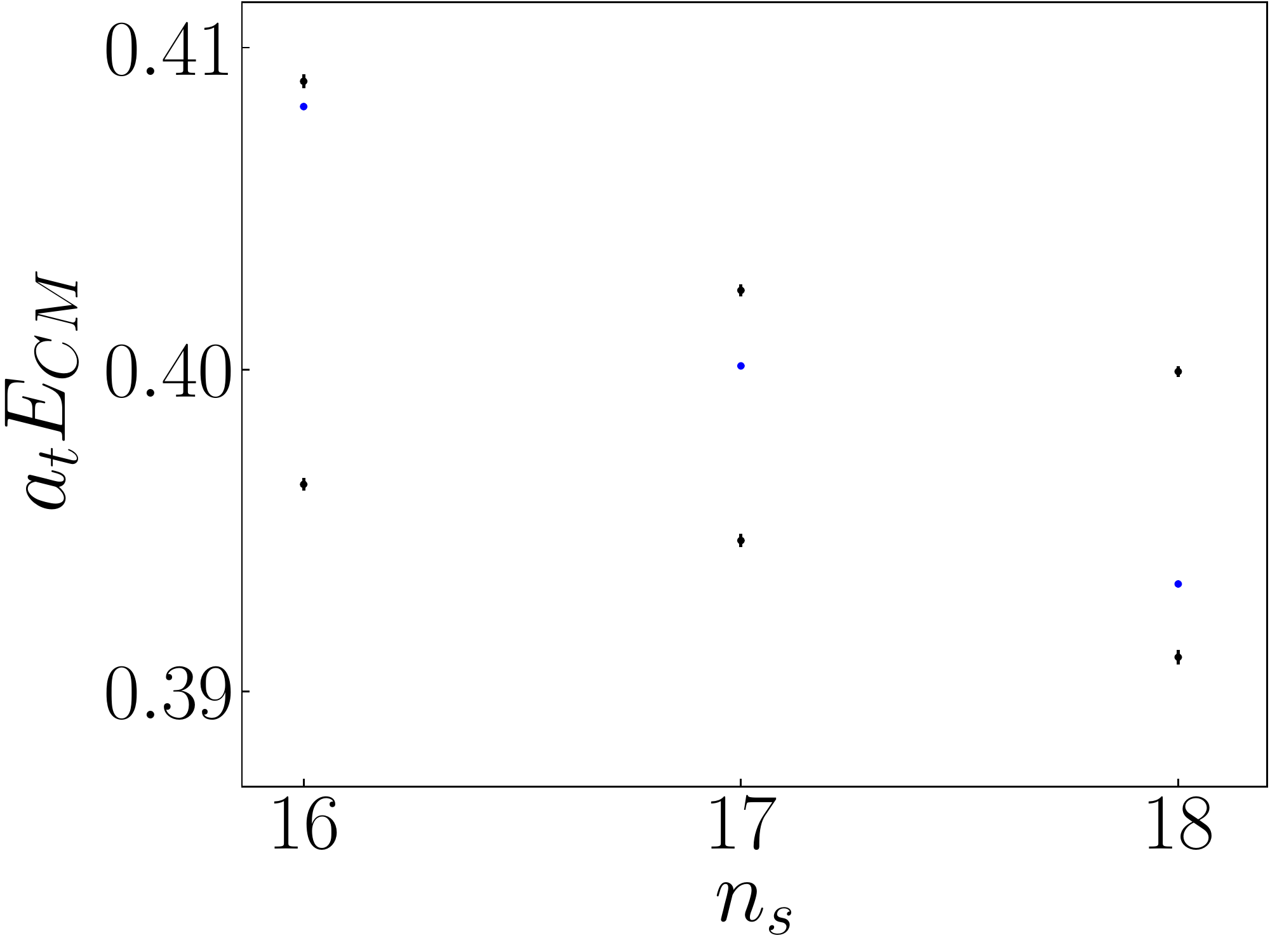}
	\end{subfigure}
	\begin{subfigure}[h]{0.49\textwidth}
		\centering
		\includegraphics[height=2in]{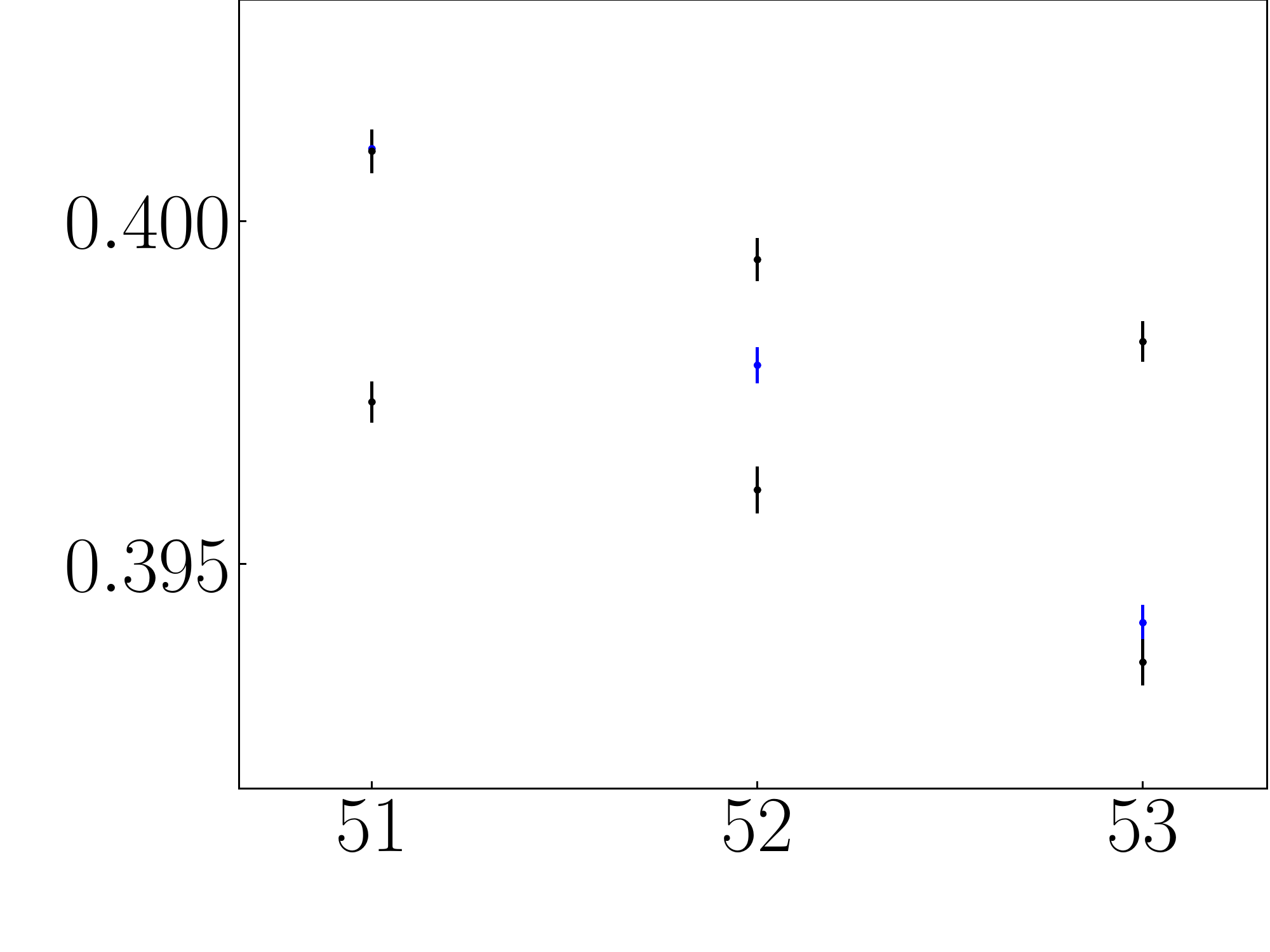}
	\end{subfigure}
\caption{Avoided level crossings in the $\vb{d}^2=1$ channel, boosted to the center-of-momentum frame.}
\end{figure}

Using the box matrix formalism developed in Ref.~\cite{Morningstar2017} and applied in works 
such as~\cite{Andersen2018, Andersen2018_2, Brett2018}, we write the quantization condition which relates
 our finite-volume spectrum to the infinite-volume $K$-matrix as,
\beq \label{eq:kb}
\det \left[\tilde K^{-1}(E_{\rm{CM}}) - B^{\Lambda,\vb d}(E_{\rm{CM}})\right] = 0,
\eeq
where $E_{CM}$ refers to a particular lab-frame energy boosted to the center-of-momentum frame. $\tilde K$ 
is related to the standard $K$-matrix as described below.  The so-called box matrix $B$ is Hermitian
and block-diagonal in the octahedral irrep, $\Lambda$, and in total momentum class, $\vb d$. For spinless 
scattering, $\tilde K$ may be indexed by only angular 
momentum $l$, but must be truncated by some maximum $l_{\rm{max}}$. In this work, we consider $l_{\rm{max}}=0$ 
and investigate the systematic errors introduced by this truncation by also considering $l_{\rm{max}}=2$. 
Note that $l=1$ is excluded by parity. It should be noted that in Eq.~(\ref{eq:kb}), each energy 
determination provides a single relation between that energy and the entire $\tilde K$-matrix. When 
truncating $\tilde K$ down to a single angular momentum, calculating $B(E_{\rm{CM}})$ (using freely 
developed software made available in Ref.~\cite{Morningstar2017}) tells us $\tilde K^{-1}$ 
directly. For higher-$l$ truncations, however, we must parametrize $\tilde K$ 
and use many calculations of the energies and $B(E_{\rm{CM}})$ to fit to the entire $\tilde K$-matrix.

Define the following center-of-mass frame kinematic variables:
\begin{equation}
E_{\rm{CM}} = \sqrt{E^2-\vb{P_{\rm{tot}}}^2}, \qquad q_{\rm{CM}} = 
\sqrt{\left(\frac{E_{\rm{CM}}}{2}\right)^2 - m_\phi^2},
\end{equation}
where $E$ is an energy measured in the lab-frame, $\vb{P_{\rm{tot}}}=\frac{2\pi\vb{d}}{L}$ is the total 
lab-frame momentum associated with that energy, and $m_\phi$ is the $\phi$ mass measured in the 
$\vb{P_{\rm{tot}}}=0$ channel. With these definitions, we can now define $\tilde K^{-1}$ in terms of 
the usual $K$-matrix and scattering phase-shift $\delta_l$,
\beq
\tilde K^{-1}_l = \left(\frac{q_{\rm{CM}}}{m_\phi}\right)^{2l+1} K^{-1}_l 
= \left(\frac{q_{\rm{CM}}}{m_\phi}\right)^{2l+1}\cot\delta_l(E_{\rm{CM}}).
\eeq
We choose to parametrize $\tilde K^{-1}_{l=0}$ by a Breit-Wigner,
\beq
\tilde K^{-1}_0 = \frac{1}{2}\sqrt{\frac{E_{\rm{CM}}^2}{m_\phi^2}-4}\left(
\frac { m _ { \rho } ^ { 2 } - E _ { C M } ^ { 2 } } { m _ { \rho } \Gamma _ { \rho } }\right),
\eeq
where the resonance mass $m_\rho$ and its decay width $\Gamma_\rho$ are parameters we fit to. $\Gamma_\rho$ is related to the tri-field coupling as $\Gamma_\rho=\frac{g^2}{32\pi m_\rho^2}\sqrt{m_\rho^2-4m_\phi^2}$.
When we include $l=2$ in our truncation, we choose to parametrize it by the first term in an effective 
range expansion,
\beq
\tilde { K } _ { 2 } ^ { - 1 } = - \frac { 1 } { m _ { \phi } ^ { 5 } a _ { 2 } }.
\eeq

\begin{figure}
\centering
\includegraphics[scale=0.5]{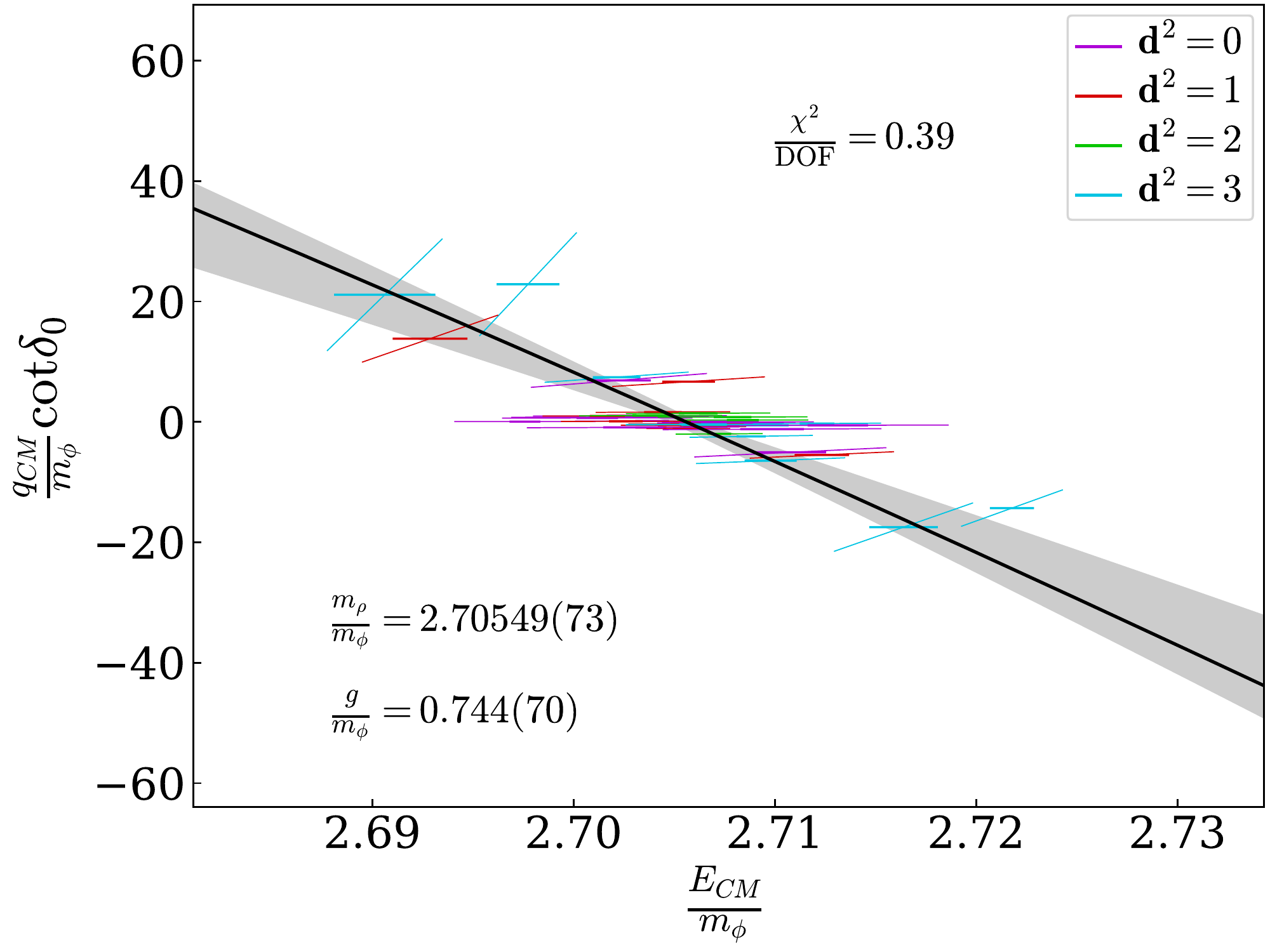}
\caption{Fit to $\tilde K^{-1}$-matrix, truncated to $l_{\rm{max}}=0$.}\label{fig:ps_fit_0}
\end{figure}

\begin{figure}
\centering
\includegraphics[scale=0.5]{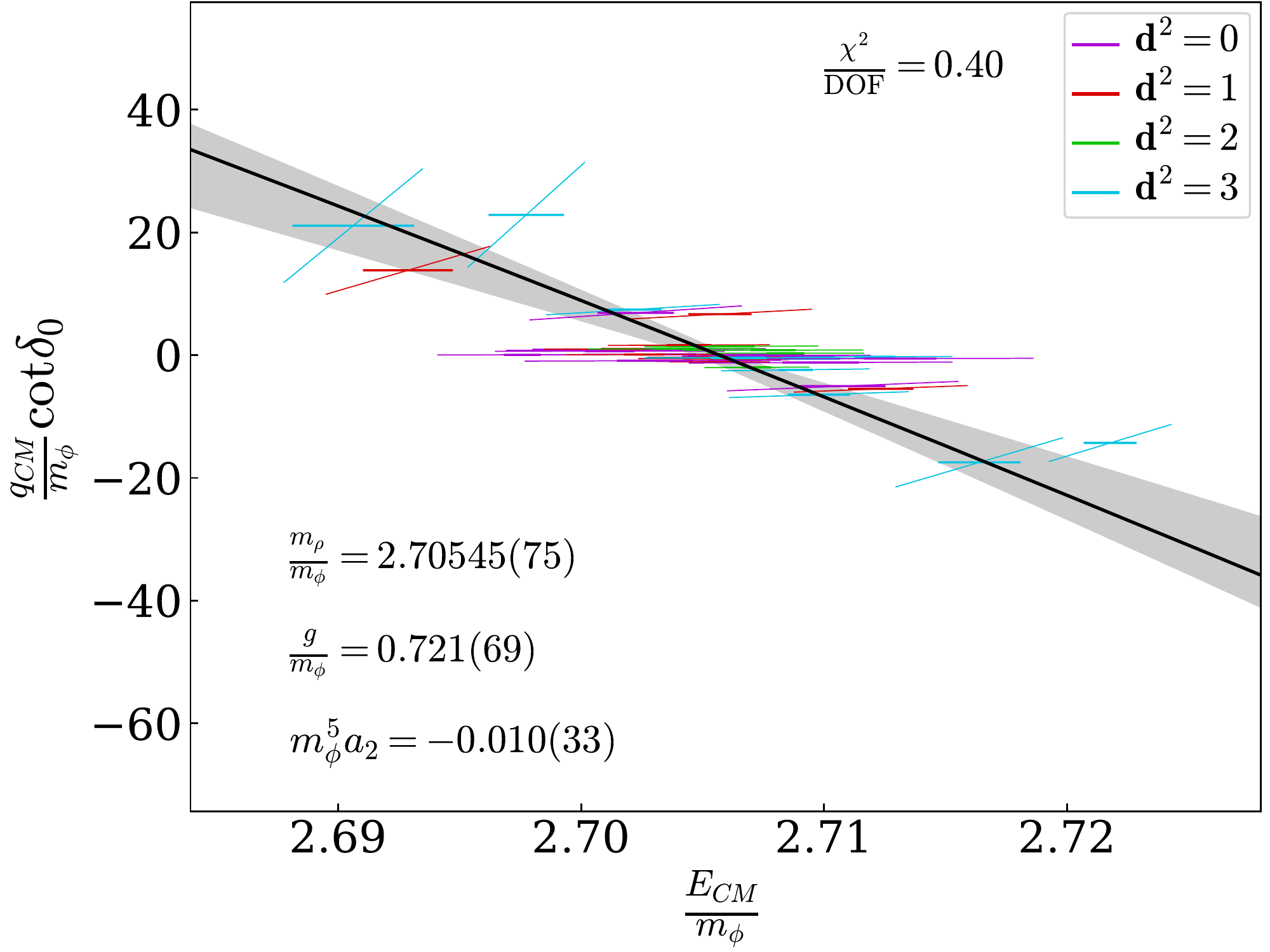}
\caption{Fit to $\tilde K^{-1}$-matrix, truncated to $l_{\rm{max}}=2$.}\label{fig:ps_fit_2}
\end{figure}

Fig.~\ref{fig:ps_fit_0} shows calculations of $\frac{q_{\rm{CM}}}{m_\phi}\cot\delta_0$ along with a curve based on a fit to an 
$l_{\rm{max}}=0$ $\tilde K$-matrix. Fig.~\ref{fig:ps_fit_2} shows a similar fit, but this time 
including the $l=2$ partial wave in our fit.
We find that not only is the $l=2$ scattering length within error of zero, but also that including the $l=2$ 
partial wave in our fit does not significantly change the determination of the resonance 
parameters $m_\rho$ and $\Gamma_\rho$. We therefore determine that it is justified to truncate 
our $\tilde K$-matrix to $l_{\rm{max}}=0$.  Our determination of the fit parameters is still
ongoing and will be published in the near future.

\acknowledgments\noindent
This work was supported by the U.S.~National Science Foundation 
under award PHY-1613449.  


\end{document}